\documentstyle[tighten,preprint,eqsecnum,aps,prd]{revtex}
\begin{document}
\draft

\hyphenation{
mani-fold
mani-folds
}


\def\Bbb{\bf}

\def\BbbR{{\Bbb R}}
\def\BbbZ{{\Bbb Z}}

\def\sltwor{{{\rm SL}(2,\BbbR)}}
\def\sltwoz{{{\rm SL}(2,\BbbZ)}}
\def\psltwoz{{{\rm PSL}(2,\BbbZ)}}

\def\suoneone{{{\rm SU}(1,1)}}

\def\fracpart{\mathop{\rm frac}\nolimits}

\def\teich{Teichm\"uller}

\def\repsym{{\sf T}}
\def\hilbertspace{{\cal H}}
\def\rinftyrep{{\Omega}}


\preprint{\vbox{\baselineskip=12pt
\rightline{Freiburg THEP--95/7}
\rightline{WISC--MILW--95--TH--13}
\rightline{gr-qc/9504035}}}
\title{Diffeomorphism invariant subspaces in
Witten's 2+1 quantum gravity
on $\BbbR\times T^2$}
\author{Domenico Giulini\cite{nico}}
\address{
Fakult\"at f\"ur Physik,
Universit\"at Freiburg,
\\
Hermann-Herder-Strasse~3,
D-79104 Freiburg, Germany}
\author{Jorma Louko\cite{jorma}}
\address{
Department of Physics,
University of
Wisconsin--Milwaukee,
\\
P.O.\ Box 413,
Milwaukee, Wisconsin 53201, USA}
\date{April 1995}
\maketitle
\begin{abstract}%
We address the role of large diffeomorphisms in Witten's 2+1
gravity on the manifold $\BbbR\times T^2$. In a ``spacelike sector"
quantum theory that treats the large diffeomorphisms as a symmetry,
rather than as gauge, the Hilbert space is shown to contain no
nontrivial finite dimensional subspaces that are invariant under the
large diffeomorphisms. Infinite dimensional closed invariant
subspaces are explicitly constructed, and the representation of the
large diffeomorphisms is thus shown to be reducible.
Comparison is made to Witten's theory on $\BbbR\times \Sigma$,
where $\Sigma$ is a higher genus surface.
\end{abstract}

\narrowtext

\section{Introduction}
\label{sec:intro}

In Witten's 2+1 gravity\cite{achu,witten1} on spacetimes of the form
${\cal M} = \BbbR \times \Sigma$, where $\Sigma$ is a closed
orientable two-manifold of genus~$g$, the simplest nontrivial case of
$g=1$ is known to differ in several qualitative ways from the generic
case of $g>1$\cite{mess,carlip1,carlip2,carlip3,louma,carlip-water}.
One facet of this is the way that large diffeomorphisms
({\it i.e\/}., diffeomorphisms disconnected from the identity)
appear in the theory.
For $g>1$, the geometrodynamically relevant connected component of
the classical solution space is the cotangent bundle over the \teich\
space $T^g$ of~$\Sigma$. The quotient group $G$ of all
diffeomorphisms modulo diffeomorphisms connected to the identity is
the modular group, and the quotient space $T^g/G$ is the Riemann
moduli space, which is a smooth manifold everywhere except at
isolated singularities. Upon quantization, one option is to take the
Hilbert space to be $L^2(T^g)$ with respect to a natural volume
element on $T^g$\cite{witten1,goldman1,goldman2,AAbook2}, and to let
the large diffeomorphisms act on this space as symmetries. Another
option is to treat the large diffeomorphisms as gauge, in which case
they should be factored out from the quantum theory; this can be
achieved by taking the Hilbert space to be~$L^2(T^g/G)$.

For $g=1$ the classical solution space is not a manifold, nor is
the subset that corresponds to conventional
geometrodynamics\cite{carlip1,louma,moncrief1}. The attention has
therefore often been fixed to the so-called ``spacelike sector" of
the theory, where the classical solution space consists of two copies
of the ``square root" geometrodynamical
theory\cite{moncrief1,hosoya-nakao1} glued together by a
lower-dimensional non-geometrodynamical part\cite{louma}. The
configuration space of this sector can be regarded as ${\cal N}_{\rm
S}:=\left(\BbbR^2\setminus\{(0,0)\}\right)/\BbbZ_2$, where the
$\BbbZ_2$ action on $\BbbR^2\setminus\{(0,0)\}$ is generated by the
map $(x^1,x^2)\mapsto(-x^1,-x^2)$, and the phase space is the
cotangent bundle over~${\cal N}_{\rm S}$. ${\cal N}_{\rm S}$ is
equipped with the volume element $d\mu := dx^1
dx^2$\cite{AAbook2,five-a}. The modular group is now $\sltwoz$, and
its action on ${\cal N}_{\rm S}$ is induced from the action on
$\BbbR^2$ given by
\begin{equation}
\left(
\begin{array}{c}
x^1 \\
x^2 \\
\end{array}
\right)
\longmapsto
M
\left(
\begin{array}{c}
x^1 \\
x^2 \\
\end{array}
\right)
\ \ , \ \
M \in \sltwoz
\ \ ,
\end{equation}
where $M$ on the right hand side acts on the column vector by usual
matrix multiplication. (Clearly, this $\sltwoz$ action on ${\cal
N}_{\rm S}$ reduces to an action of the factor group
$\psltwoz=\sltwoz/\{\openone,-\openone\}$, where $\openone$ stands
for the two by two unit matrix.) If now the large diffeomorphisms are
understood as a symmetry, the Hilbert space can be taken to be
${\hilbertspace}_{\rm S} := L^2({\cal N}_{\rm S};
d\mu)$\cite{carlip1,louma,AAbook2,five-a}, and the action of
$\sltwoz$ on ${\cal N}_{\rm S}$ clearly induces a unitary
representation $\repsym^{\rm S}_\sltwoz$ of $\sltwoz$
on~${\hilbertspace}_{\rm S}$. Treating the large diffeomorphisms as
gauge is more problematic, however. One attempt might be to follow
the logic of the higher genus surfaces and regard the quotient space
${\cal N}_{\rm S}/\sltwoz$ as a configuration space on which the
quantum theory is to be built. However, the action of $\sltwoz$ on
${\cal N}_{\rm S}$ is not properly discontinuous. In fact, each
half-line with rational $x^2/x^1$ is fixed by an infinite Abelian
subgroup, whereas the half-lines with irrational $x^2/x^1$ are fixed
only by $\pm\openone$\cite{peldan}. This implies that the isomorphism
class of stabilizer subgroups is nowhere locally constant. The
quotient space ${\cal N}_{\rm S}/\sltwoz$ is thus nowhere locally a
manifold, and it is not obvious whether one could use such a space as
a configuration space for the quantum theory. An alternative attempt
might be to employ the Hilbert space~${\cal H}_{\rm S}$, but to
restrict observables to those that commute with the given unitary
action of $\sltwoz$. This prevents observables from having non-zero
matrix elements between states belonging to inequivalent
subrepresentations. Therefore, no state that can be decomposed into
the sum of states belonging to inequivalent subrepresentations can be
a pure state. Here, ``state" is understood to mean ``state for the
algebra of observables;" see {\it e.g\/}.\cite{bogolubov}. However,
from the results in Sections \ref{sec:sltwor} and \ref{sec:sltwoz} it
follows that {\it every\/} state can be written in such a form (as
will be discussed more explicitly in the following paragraph). The
attempt to reduce the gauge redundancy in this fashion thus leads to
the somewhat paradoxical result that there are no pure states.

That the gauge interpretation of large diffeomorphisms leads to
the absence of pure states is, in fact, not at all
surprising. This can be seen from a more familiar example:
Consider the Hilbert space $\hilbertspace=L^2(\BbbR^2; dx^1dx^2)$
and the unitary action of the translation group in
$x^1$-direction. We want to regard the translations
(for simplicity we shall refer to the $x^1$-translations
simply as translations) as gauge, and
hence require all  observables to commute with them. Their integral
kernels, $O(x^1,x^2;y^1,y^2)$, thus only depend on $x^1-y^1$, $x^2$,
and~$y^2$. On the other hand, each translation invariant subspace,
$\hilbertspace_{\Delta}\subset {\cal H}$, is uniquely given
by those functions whose Fourier transform in $x^1$ vanishes
almost everywhere outside the measurable set
$\Delta\subset \BbbR$ in the transformed $x^1$
coordinate\cite{rudin}. Given two disjoint measurable sets,
$\Delta_1$ and~$\Delta_2$, it is easy to see that all matrix
elements of observables between states in
${\hilbertspace}_{\Delta_1}$ and $\hilbertspace_{\Delta_2}$ vanish.
This is a direct consequence of the fact that translation invariant
operators cannot increase the support of the Fourier transform,
and can also be checked by direct calculation using the
property of integral kernels given above. On the other hand, given a
measurable set $\Delta$ of non-zero measure, it is always possible to
find a measurable subset $\Delta_1\subset\Delta$
of non-zero but strictly smaller measure.
(A~short proof of this fact will be given in
Appendix~\ref{app:lebesgue}.) Denoting by $\Delta_2$ the complement
of $\Delta_1$ in~$\Delta$, we have a decomposition
$\Delta=\Delta_1\cup\Delta_2$ of $\Delta$ into disjoint subsets of
non-zero measures. If we take for $\Delta$ the support in the Fourier
transformed $x^1$ coordinate of an arbitrary element
in~$\hilbertspace$, we immediately infer that {\it any\/} vector is
the sum of two vectors which lie in different and strictly smaller
invariant subspaces. Hence there are no pure states. A~similar
argument applies to our gravitational case, using the diffeomorphism
invariant subspaces of Theorem \ref{sec:sltwor}.2 in
Section~\ref{sec:sltwor}.

In the above simple example it is clear how the redundant
translations are properly eliminated: instead of $\hilbertspace$ one
considers the Hilbert space $L^2(\BbbR;dx^2)$ of square integrable
functions over the classical reduced configuration space, which may
be identified with the $x^2$ axis. However, in our gravitational case
the analogous option is not at our direct disposal, due to the
complicated structure of ${\cal N}_{\rm S}/\sltwoz$. In this paper we
shall therefore concentrate on the theory in which the large
diffeomorphisms are treated as symmetries. The Hilbert space is
thus~$\hilbertspace_{\rm S}$, the large diffeomorphisms act on
$\hilbertspace_{\rm S}$ by $\repsym^{\rm S}_\sltwoz$, and the algebra
of observables is taken to be the full algebra $B(\hilbertspace_{\rm
S})$ of bounded operators on~$\hilbertspace_{\rm S}$. Note that this
means allowing, in principle, observables that do not commute with
the large diffeomorphisms.  At the fundamental level this theory has
no superselection rules, and rays in $\hilbertspace_{\rm S}$ are in
bijective correspondence to pure states.

The purpose of this paper is to make two observations about the
unitary representation $\repsym^{\rm S}_\sltwoz$ of $\sltwoz$ on
${\hilbertspace}_{\rm S}$. On the one hand, we point out that
$\repsym^{\rm S}_\sltwoz$ is reducible, and we exhibit a class of
infinite dimensional closed invariant subspaces. On the other hand,
we demonstrate that ${\hilbertspace}_{\rm S}$ contains no nontrivial
finite dimensional invariant subspaces.\footnote{While this paper was
in preparation, an independent argument ruling out nontrivial finite
dimensional invariant subspaces was given in the revised version of
Ref.\cite{peldan}. We thank Peter Peld\'an for discussions on this
issue.}

Our starting point is the decomposition\cite{knapp-prob,howe-tan-ex}
of the standard unitary representation $\repsym_\sltwor$ of $\sltwor$
on $L^2(\BbbR^2)$ into a direct integral of irreducible unitary
representations, all of whom belong to the principal
series\cite{bargmann,lang,knapp,howe-tan}. This decomposition yields
an obvious construction of closed infinite dimensional subspaces of
$L^2(\BbbR^2)$ that are invariant under~$\repsym_\sltwor$. Projecting
$L^2(\BbbR^2)$ to the subspace~${\hilbertspace}_{\rm S}$ then
produces closed infinite dimensional subspaces of
${\hilbertspace}_{\rm S}$ that are invariant under $\repsym_\sltwor$,
and hence also under $\repsym^{\rm S}_\sltwoz$. This is the first
claim above.

To prove the second claim, let us denote by $\repsym_\sltwoz$ the
restriction of $\repsym_\sltwor$ to $\sltwoz$. It is known that the
principal series irreducible unitary representations of $\sltwor$
restrict to irreducible representations of $\sltwoz$\cite{cowsteg}.
(Further, two representations of $\sltwoz$ obtained in this fashion
are equivalent only if the corresponding representations of $\sltwor$
are\cite{cowsteg,bishsteg}.) This means that the direct integral
decomposition of $\repsym_\sltwor$ yields, through restriction to
$\sltwoz$, a decomposition of $\repsym_\sltwoz$ into a direct
integral of irreducible infinite dimensional representations. It
follows immediately that $L^2(\BbbR^2)$ has no nontrivial finite
dimensional subspaces that are invariant under~$\repsym_\sltwoz$.
This implies the claim.

With the decomposition of~$\repsym_\sltwoz$, one can translate the
action of $\sltwoz$ on the configuration space ${\cal N}_{\rm S}$
into an action of $\sltwoz$ on~$S^1$, where the $S^1$ arises as the
configuration space of the constituent irreducible representations of
$\sltwoz$ on~$L^2(S^1)$. The problem of building a quantum theory
with the configuration space ${\cal N}_{\rm S}/\sltwoz$ is thus
translated into the problem of building quantum theories with the
configuration space $S^1/\sltwoz$. We show that the difficulty
persists: the action of $\sltwoz$ on $S^1$ is not properly
discontinuous, and $S^1/\sltwoz$ is nowhere locally a manifold.

The rest of the paper is as follows. In Section \ref{sec:sltwor} we
present the decomposition of the standard representation of $\sltwor$
on $L^2(\BbbR^2)$ into a direct integral of irreducible
representations, and we use this decomposition to construct a class
of closed invariant subspaces. The restriction to  the subgroup
$\sltwoz$ is addressed in Section~\ref{sec:sltwoz}. Section
\ref{sec:discussion} contains a brief discussion. Appendix
\ref{app:lebesgue} recalls an elementary property of the Lebesgue
measure, and Appendix \ref{app:Sone} contains an analysis of the
quotient space $S^1/\sltwoz$.

\section{Representation of $\sltwor$ on $L^2(\BbbR^2)$}
\label{sec:sltwor}

In this section we first review the decomposition of the standard
unitary representation of $\sltwor$ on $L^2(\BbbR^2)$ into
irreducible unitary representations\cite{knapp-prob}. We then note
that the decomposition presents an obvious way of constructing
infinite dimensional closed invariant subspaces of~$L^2(\BbbR^2)$.

Let $(x^1,x^2)$ be a pair of global coordinates on~$\BbbR^2$.
The group $\sltwor$ has on $\BbbR^2$ the natural associative
action
\begin{equation}
\left(
\begin{array}{c}
x^1 \\
x^2 \\
\end{array}
\right)
\longmapsto
M
\left(
\begin{array}{c}
x^1 \\
x^2 \\
\end{array}
\right)
\ \ , \ \
M \in \sltwor
\ \ ,
\label{sltworactionR}
\end{equation}
where $M$ on the right hand side acts on the column vector by usual
matrix multiplication. Denoting a point on $\BbbR^2$ by~$x$, we write
this action as as $x\mapsto Mx$.

Let ${\hilbertspace} := L^2(\BbbR^2)=L^2(\BbbR^2;dx^1dx^2)$ be the
Hilbert space of square integrable
functions\footnote{Note that functions represent the same element in
$L^2$ spaces if they differ at most on a set of measure zero.
We shall therefore throughout understand functions to be defined
only almost everywhere~(a.e.), and  pointwise equations for the
functions to hold only a.e.}
on $\BbbR^2$ with the inner product
\begin{equation}
(f,g) := \int dx^1 dx^2 \, {\overline f}g
\ \ .
\label{ip}
\end{equation}
We define a representation $\repsym$ of $\sltwor$
on~$\hilbertspace$,
$f\mapsto \repsym(M)f$, by
\begin{equation}
\repsym(M)f(x) := f (M^{-1}x)
\ \ .
\label{sltworactionH}
\end{equation}
This representation is clearly unitary,
$(\repsym(M)f,\repsym(M)g)=(f,g)$. Our aim is to decompose this
representation into its irreducible components. As $\sltwor$ is a
Type~I group, the decomposition is essentially unique\cite{mackey}.

We first rewrite the $\sltwor$ action on $\BbbR^2$
(\ref{sltworactionR}) in a more convenient manner. This action
clearly leaves the origin invariant.
On $\BbbR^2\setminus\{(0,0)\}$,
introduce the polar coordinates $(r,\theta)$ through
\begin{equation}
\begin{array}{rl}
&x^1 = r \cos\theta
\ \ ,
\\
&x^2 = r \sin\theta
\ \ ,
\label{polarcoords}
\end{array}
\end{equation}
where $r>0$, and $\theta$ is understood periodic with period~$2\pi$.
We parametrize a matrix $M\in\sltwor$ as
\begin{equation}
M = U
\left(
\begin{array}{rr}
\alpha & \beta \\
{\bar\beta} & {\bar\alpha} \\
\end{array}
\right)
U^{-1}
\ \ ,
\label{suoneonepar}
\end{equation}
where $U$ is the unitary matrix
\begin{equation}
U :=
{1 \over \sqrt{2}}
\left(
\begin{array}{rr}
1 & 1 \\
i & -i \\
\end{array}
\right)
\ \ ,
\end{equation}
and $\alpha$ and $\beta$ are complex numbers satisfying
$\alpha{\bar\alpha}-\beta{\bar\beta}=1$. This parametrization is
one-to-one, and if the matrix
$\pmatrix{ \alpha & \beta \cr
{\bar\beta} & {\bar\alpha} \cr}$
is interpreted as an element of $\suoneone$,
(\ref{suoneonepar}) defines an isomorphism
$\sltwor \simeq \suoneone$\cite{bargmann}.
The $\sltwor$ action (\ref{sltworactionR}) on
$\BbbR^2\setminus\{(0,0)\}$ takes then the form
$(r,\theta) \mapsto (Mr,M\theta)$, where
\begin{mathletters}
\label{sltwoactionpolar}
\begin{eqnarray}
e^{iM\theta} &=& e^{i\theta}
{{\overline{W(M,\theta)}}
\over
|W(M,\theta)|}
\ \ ,
\label{sltwoactiontheta}
\\
\noalign{\bigskip}
Mr &=& r |W(M,\theta)|
\ \ ,
\end{eqnarray}
\end{mathletters}
with
\begin{equation}
W(M,\theta) := \alpha + \beta e^{2i\theta}
\ \ .
\end{equation}

Next, we perform a radial Mellin transform on~${\hilbertspace}$. For
$f\in{\hilbertspace}$, its transform ${\hat f}$ is defined by
\begin{equation}
{\hat f} (s, \theta)
:=
\int_0^\infty dr \, r^{is} f(r,\theta)
\ \ , \ \
s\in\BbbR
\ \ .
\label{mellin}
\end{equation}
Here, and from now on, we understand the argument of a function in
${\hilbertspace}$ to be the pair of polar coordinates
$(r,\theta)$~(\ref{polarcoords}). The transform (\ref{mellin})
defines an isomorphism
${\hilbertspace}\simeq {\hat{\hilbertspace}}
:= L^2(\BbbR
\times S^1; {(2\pi)}^{-1} ds d\theta)$:
the inner product (\ref{ip}) can be written as
\begin{equation}
(f,g) =
{1\over 2\pi} \int_{-\infty}^\infty ds
\int_{-\pi}^\pi d\theta \,
{\overline{{\hat f} (s,\theta)}} \,
{\hat g} (s,\theta)
\ \ ,
\label{ip1}
\end{equation}
and the inverse transform is
\begin{equation}
f(r,\theta) =
{1\over 2\pi} \int_{-\infty}^\infty ds \,
r^{-1-is} \, {\hat f} (s, \theta)
\ \ .
\label{imellin}
\end{equation}
These statements follow directly from the observation that in terms
of the logarithmic radial coordinate $t=\ln r$, the inner product
(\ref{ip}) reads \begin{equation}
(f,g) =
\int_{-\infty}^\infty dt
\int_{-\pi}^\pi d\theta \,
{\overline{e^t f(e^t,\theta)}} \,
{e^t g(e^t,\theta)}
\ \ ,
\end{equation}
and the transforms (\ref{mellin}) and (\ref{imellin}) reduce to an
ordinary Fourier transform pair,
\begin{mathletters}
\begin{eqnarray}
{\hat f} (s, \theta)
&=&
\int_{-\infty}^\infty dt \,
e^{ist} \, e^t f(e^t,\theta)
\ \ ,
\\
\noalign{\smallskip}
e^t f(e^t,\theta)
&=&
{1\over 2\pi} \int_{-\infty}^\infty ds \,
e^{-ist} {\hat f} (s, \theta)
\ \ .
\end{eqnarray}
\end{mathletters}

By (\ref{sltwoactionpolar}) and~(\ref{mellin}), the Mellin transform
maps the representation $\repsym$ of $\sltwor$ into a
representation ${\hat \repsym}$ on~${\hat{\hilbertspace}}$,
given by
\begin{equation}
{\hat \repsym}(M){\hat f} (s,\theta)
=
{| W(M, M^{-1} \theta)|}^{1+is}
{\hat f} (s, M^{-1}\theta)
\ \ .
\label{sltworactionHhat}
\end{equation}
The remarkable property of (\ref{sltworactionHhat}) is that the
different values of $s$ are decoupled. To utilize this, we write
${\hat{\hilbertspace}}$ as the direct integral
\begin{equation}
{\hat{\hilbertspace}}
= \int_{-\infty}^\infty ds \,
{\hat{\hilbertspace}}_s
\ \ ,
\end{equation}
where ${\hat{\hilbertspace}}_s
\simeq L^2 (S^1; {(2\pi)}^{-1} d\theta)$
with the inner product
\begin{equation}
({\hat f}_s,{\hat g}_s)_s =
{1\over 2\pi} \int_{-\pi}^\pi d\theta \,
{\overline{{\hat f}_s (\theta)}} \,
{\hat g}_s (\theta)
\ \ , \ \
f_s, g_s \in {\hat{\hilbertspace}}_s
\ \ .
\label{ips}
\end{equation}
The representation ${\hat \repsym}$ (\ref{sltworactionHhat}) then
decomposes into a representation ${\hat \repsym}_s$ on
each~${\hat{\hilbertspace}}_s$, given by
\begin{equation}
{\hat \repsym}_s(M){\hat f}_s (\theta) =
{| W(M, M^{-1} \theta)|}^{1+is}
{\hat f}_s (M^{-1}\theta)
\ \ .
\label{sltworactionHhats}
\end{equation}
It is straightforward to verify that ${\hat \repsym}_s$ is a unitary
representation for every~$s$. However, it is not
irreducible.

To proceed, we write
${\hat{\hilbertspace}}_s
={\hat{\hilbertspace}}_s^+
\oplus {\hat{\hilbertspace}}_s^-$,
where ${\hat{\hilbertspace}}_s^\pm$ are the two closed subspaces of
${\hat{\hilbertspace}}_s$ where the functions satisfy respectively
${\hat f}_s(\theta+\pi)=\pm{\hat f}_s(\theta)$.
${\hat{\hilbertspace}}_s^\pm$ are clearly each invariant
under~${\hat\repsym}_s$.
Therefore, ${\hat\repsym}_s$ decomposes
into unitary representations of $\sltwor$
on~${\hat{\hilbertspace}}_s^\pm$. We shall show that, with the
exception of~${\hat{\hilbertspace}}_0^-$, all these representations
are irreducible.

Let $s$ be fixed, and consider~${\hat{\hilbertspace}}_s^+$. For
${\hat f}\in{\hat{\hilbertspace}}_s^+$,
we can write ${\hat f}(\theta) =
{\tilde f}(2\theta)$,
where ${\tilde f}$ is periodic in its argument with
period~$2\pi$. (For brevity, we drop the subscript $s$ when there is
no danger of confusion.) This gives an isomorphism
${\hat{\hilbertspace}}_s^+
\simeq {\tilde{\hilbertspace}}_s^+
:= L^2(S^1; {(2\pi)}^{-1}d\phi)$,
where the inner product induced from (\ref{ips}) is
\begin{equation}
({\tilde f},{\tilde g})_s
=
{1\over 2\pi} \int_{-\pi}^\pi d\phi \,
{\overline{{\tilde f} (\phi)}} \,
{\tilde g} (\phi)
\ \ .
\label{ipts+}
\end{equation}
The resulting representation ${\tilde \repsym}_s^+$ of $\sltwor$
on ${\tilde{\hilbertspace}}^+$ is then
given by
\begin{equation}
{\tilde \repsym}_s^+(M){\tilde f} (\phi)
=
{| w(M, M^{-1} \phi)|}^{1+is}
{\tilde f} (M^{-1}\phi)
\ \ ,
\label{sltworactionHt+}
\end{equation}
where
\begin{equation}
w(M, \phi) := \alpha + \beta e^{i\phi}
\ \ ,
\end{equation}
and the action of $\sltwor$ on $\phi$ is determined by
(\ref{sltwoactiontheta}) and takes the form
\begin{equation}
e^{iM\phi} = e^{i\phi}
{{\overline{w(M,\phi)}}
\over
w(M,\phi)}
\ \ .
\label{sltwoactionphi}
\end{equation}
This is recognized as the irreducible unitary representation of the
continuous class $C^0_q$ constructed in
Ref.\cite{bargmann}\footnote{Note that formula (6.11) in
Ref.\cite{bargmann} has a
typographical error and should read
$T_\sigma(a)f(\phi) =
\mu(a, a^{-1}\phi)^{{1\over2}+\sigma} f(a^{-1}\phi)$.},
with the Casimir invariant $q$ taking the value $(1+s^2)/4$.
These representations are known as the principal
series of even parity\cite{lang,knapp,howe-tan}.

Let then $s$ again be fixed, and
consider~${\hat{\hilbertspace}}_s^-$. For
${\hat f}\in{\hat{\hilbertspace}}_s^-$, we now can write
${\hat f}(\theta) =
e^{i\theta} {\tilde f}(2\theta)$, where ${\tilde f}$ is periodic in
its argument with period~$2\pi$. This gives an isomorphism
${\hat{\hilbertspace}}_s^-\simeq {\tilde{\hilbertspace}}_s^- :=
L^2(S^1; {(2\pi)}^{-1}d\phi)$,
where the inner product induced from
(\ref{ips}) is again given by~(\ref{ipts+}). The resulting
representation ${\tilde \repsym}_s^-$ of
$\sltwor$ on ${\tilde{\hilbertspace}}^-$ is then given by
\begin{equation}
{\tilde \repsym}_s^-(M){\tilde f} (\phi)
=
{| w(M, M^{-1} \phi)|}^{1+is}
\nu(M, M^{-1}\phi)
{\tilde f} (M^{-1}\phi)
\ \ ,
\label{sltworactionHt-}
\end{equation}
where
\begin{equation}
\nu(M,\phi) :=
{w(M, \phi)
\over
|w(M, \phi)|
}
\ \ ,
\end{equation}
and the rest of the notation is as with~${\tilde \repsym}_s^+$. For
$s\ne0$, ${\tilde \repsym}_s^-$ is recognized as the irreducible
unitary representation of the continuous class $C^{1/2}_q$
constructed in Ref.\cite{bargmann},
with the Casimir invariant $q$ taking the value $(1+s^2)/4$.
These representations are known as the principal series of odd
parity\cite{lang,knapp,howe-tan}.
The representation ${\tilde \repsym}_0^-$ decomposes
into a direct sum of two irreducible unitary representations, denoted
in Ref.\cite{bargmann} by $D^+_{1/2}$ and $D^-_{1/2}$ and known as
the limits of the discrete series\cite{lang,knapp,howe-tan}.

We thus have a complete decomposition of the representation $\repsym$
(\ref{sltworactionH}) of $\sltwor$ on~${\hilbertspace}$ into its
irreducible components. We collect the statements into a
theorem\cite{knapp-prob}.

{\bf Theorem \ref{sec:sltwor}.1.}
The Hilbert space ${\hilbertspace}= L^2(\BbbR^2)$ has a decomposition
\begin{equation}
{\hilbertspace} \simeq
\int\limits_{-\infty}^\infty ds \,
\left(
{\tilde{\hilbertspace}}^+_s \oplus {\tilde{\hilbertspace}}^-_s
\right)
\ \ ,
\label{H-full-decomp}
\end{equation}
where the integral is a direct integral, such that
(i) ${\tilde{\hilbertspace}}^\pm_s \simeq L^2(S^1)$ for every~$s$;
(ii) the unitary representation $\repsym$ (\ref{sltworactionH}) of
$\sltwor$ on ${\hilbertspace}$ decomposes into the unitary
representations~${\tilde\repsym}^\pm_s$
on~${\tilde{\hilbertspace}}^\pm_s$;
(iii) ${\tilde \repsym}^+_s$ is an irreducible unitary representation
in the principal series with even parity, $C^0_q$, with
$q=(1+s^2)/4$;
(iv) ${\tilde \repsym}^-_s$ with $s\ne0$ is an irreducible unitary
representation in the principal series with odd parity, $C^{1/2}_q$,
with $q=(1+s^2)/4$.~~$\Box$

Note that the further decomposition of~${\tilde \repsym}^-_0$ is not
relevant in Theorem \ref{sec:sltwor}.1, as the point $s=0$ is a set
of measure zero on the real line and therefore does not contribute to
the integral in~(\ref{H-full-decomp}). Note also that if we write
\begin{mathletters}
\label{hpm-sum}
\begin{eqnarray}
\hilbertspace &=& \hilbertspace^+ \oplus \hilbertspace^-
\ \ ,
\label{h-decomp}
\\
\noalign{\smallskip}
{\hat{\hilbertspace}} &=& {\hat{\hilbertspace}}^+ \oplus
{\hat{\hilbertspace}}^-
\ \ ,
\end{eqnarray}
\end{mathletters}%
where $\hilbertspace^\pm$ consist of those functions $f$ in
$\hilbertspace$ that satisfy $f(r,\theta+\pi) = \pm f(r,\theta)$,
and similarly
${\hat{\hilbertspace}}^\pm$ consist of those functions ${\hat f}$ in
${\hat{\hilbertspace}}$ that satisfy ${\hat f}(s,\theta+\pi) = \pm
{\hat f}(s,\theta)$,
we then have
\begin{equation}
\hilbertspace^\pm \simeq
{\hat\hilbertspace}^\pm
= \int\limits_{-\infty}^\infty ds \, {\hat\hilbertspace}^\pm_s
\> \simeq \int\limits_{-\infty}^\infty ds \,
{\tilde\hilbertspace}^\pm_s
\ \ .
\label{hpm-string}
\end{equation}

As $\sltwor$ has no nontrivial finite dimensional unitary
representations\cite{howe-tan}, $\hilbertspace$ cannot have
nontrivial finite dimensional subspaces that are invariant
under~$\repsym$. There are, however, infinite dimensional closed
subspaces of ${\hilbertspace}^\pm$ that are invariant
under~$\repsym$. We have the following theorem.

{\bf Theorem \ref{sec:sltwor}.2.}
Let $E$ be a measurable subset of~$\BbbR$, and let
\begin{equation}
{\hat{\hilbertspace}}^\pm_E = \left\{ \, {\hat f} \in
{\hat{\hilbertspace}}^\pm \mid \hbox{${\hat f}(s,\theta)=0$ for
a.e.\ $s\notin E$} \, \right\}
\ \ .
\end{equation}
Then ${\hat{\hilbertspace}}^\pm_E$ is a closed subspace
of~${\hat{\hilbertspace}}^\pm$, and it is invariant under~${\hat
\repsym}$.

{\sl Proof\/}. We can assume that $E$ and
$\BbbR \setminus E$ both have strictly positive measure
(since otherwise ${\hat{\hilbertspace}}^\pm_E = \{0\}$
or ${\hat{\hilbertspace}}^\pm_E = {\hat{\hilbertspace}}^\pm$).
It is clear that
${\hat{\hilbertspace}}^\pm_E$ is an invariant subspace
of~${\hat{\hilbertspace}}^\pm$. Closedness follows from the
observation that ${\hat{\hilbertspace}}^\pm_E$ is the orthogonal
complement in ${\hat{\hilbertspace}}^\pm$ of the subspace
${\hat{\hilbertspace}}^\pm_{\BbbR\setminus E}$.~~$\Box$

The construction of ${\hat{\hilbertspace}}^\pm_E$ closely parallels
the construction of closed translationally invariant subspaces
of~$L^2(\BbbR)$. For a measurable set $E\subset\BbbR$, the functions
in $L^2(\BbbR)$ whose Fourier transform vanishes almost everywhere
outside $E$ constitute a closed translationally invariant subspace
of~$L^2(\BbbR)$; conversely, every closed translationally invariant
subspace of $L^2(\BbbR)$ is of this form for some~$E$\cite{rudin}. We
shall not address here the question as to whether the spaces
${\hat{\hilbertspace}}^\pm_E$ exhaust all closed ${\hat
\repsym}$-invariant subspaces of~${\hat{\hilbertspace}}^\pm$.

\section{Representation of $\sltwoz$ on $L^2(\BbbR^2)$}
\label{sec:sltwoz}

We now consider the consequences of the above results
when $\sltwor$ is restricted to the subgroup $\sltwoz$.

It is clear that all the unitary representations of $\sltwor$
appearing in Section \ref{sec:sltwor} restrict to unitary
representations of $\sltwoz$. The spaces
${\hat{\hilbertspace}}^\pm_E$ of Theorem \ref{sec:sltwor}.2 are
therefore invariant also under the representation of $\sltwoz$ that
is inherited from~$\repsym$. In the physical terminology introduced
in Section~\ref{sec:intro}, this implies that
${\hat{\hilbertspace}}^+_E$ are closed diffeomorphism invariant
subspaces of $\hilbertspace_{\rm S} \simeq {\hat{\hilbertspace}}^+$.

To examine the possibility of finite dimensional diffeomorphism
invariant subspaces, let us denote by ${\hat\repsym}'$ and ${\tilde
\repsym}^{\prime \pm}_s$ the representations of $\sltwoz$ that are
obtained by restriction from respectively ${\hat\repsym}$
and~${\tilde \repsym}_s^\pm$. Let ${\cal F}^\pm$ be a finite
dimensional subspace of~${\hat{\hilbertspace}}^\pm$, and let $\left\{
\, {\hat f}^\pm_k \in {\hat{\hilbertspace}}^\pm \mid \hbox{$k =
1,\ldots,N$} \, \right\}$ be a finite set of vectors spanning~${\cal
F}^\pm$. Suppose now that ${\cal F}^\pm$ is invariant
under~${\hat\repsym}'$. It follows that for a.e.\ $s\in\BbbR$, the
subspace ${\cal F}_s^\pm$ of ${\hat{\hilbertspace}}_s^\pm \simeq
{\tilde{\hilbertspace}}_s^\pm$ that is spanned by the functions
$\left\{ \, {\hat f}^\pm_k(s,\theta) \, \right\}$ is invariant
under~${\tilde \repsym}^{\prime \pm}_s$. But since
${\tilde{\hilbertspace}}_s^\pm$ are infinite dimensional and ${\tilde
\repsym}^{\prime \pm}_s$ are irreducible (except for~${\tilde
\repsym}^{\prime -}_0$)\cite{cowsteg}, ${\cal F}_s^\pm$ must be
trivial for a.e.~$s$. This implies that every ${\hat f}^\pm_k$ is the
zero vector in~${\hat{\hilbertspace}}^\pm$, and hence ${\cal F}^\pm =
\{0\}$.

Therefore, $\hilbertspace$ has no nontrivial finite dimensional
subspaces that are invariant under~${\hat\repsym}'$. In the physical
terminology of Section~\ref{sec:intro}, this implies that
$\hilbertspace_{\rm S} \simeq {\hat{\hilbertspace}}^+$ has no
nontrivial finite dimensional subspaces invariant under the large
diffeomorphisms.

\section{Discussion}
\label{sec:discussion}

In this paper we have addressed the role of large diffeomorphisms in
Witten's 2+1 gravity on the manifold $\BbbR\times T^2$. We
concentrated on a ``spacelike sector" quantum theory that treats the
large diffeomorphisms as a symmetry. On the one hand, we showed that
the Hilbert space contains no nontrivial finite dimensional subspaces
that are invariant under the large diffeomorphisms. On the other
hand, we constructed explicitly a class of infinite dimensional
closed invariant subspaces. The existence of such subspaces implies,
in particular, that the representation of the large diffeomorphisms
on the Hilbert space is reducible.

These results shed light on both the similarities and differences
between the behavior of Witten's theory on the manifold $\BbbR\times
T^2$ and the manifolds $\BbbR\times \Sigma$, where $\Sigma$ is a
surface of genus $g>1$\cite{witten1,mess,carlip1,goldman2}. For
$g>1$, a geometrodynamically relevant quantum theory that treats the
large diffeomorphisms as a symmetry is obtained by taking the
configuration space to be the \teich\ space~$T^g$. The quotient of
$T^g$ under the action of the large diffeomorphisms is the Riemann
moduli space: as $T^g$ contains infinitely many copies of the Riemann
moduli space, the quantum theory has no nontrivial finite dimensional
diffeomorphism invariant subspaces. This is similar to what we have
found for the torus. On the other hand, the differences between the
torus and the higher genus surfaces manifest themselves when one
attempts to treat the large diffeomorphisms as gauge. As the Riemann
moduli space is a manifold except at isolated singularities, a higher
genus theory that treats the large diffeomorphisms are gauge can be
obtained by defining the inner product by an integral over just the
Riemann moduli space instead of all of~$T^g$. In contrast, the
corresponding quotient space for the torus seems too pathological to
be employed in a similar fashion\cite{peldan}. We shall show in
Appendix \ref{app:Sone} that this pathology persists even when the
torus Hilbert space is decomposed by (\ref{hpm-string}) into a direct
integral of Hilbert spaces that carry irreducible representations of
the large diffeomorphism group. An attempt to reduce the gauge
redundancies in the torus theory at the quantum level leads to the
absence of any pure states in the theory, as discussed in the
Introduction.

For $\BbbR\times T^2$, the construction of a connection
representation quantum theory where the large diffeomorphisms are
treated as gauge remains thus an open problem. In the metric-type
representations of
Refs.\cite{carlip1,carlip2,carlip3,carlip-water,peldan2}, the
difficulty does not appear.

Finally, it should be emphasized that we have not attempted to
interpret physically the symmetries generated by the large
diffeomorphisms. Doing so would require, among other things, a
physical interpretation of those observables that do not commute with
the large diffeomorphisms. For noncompact two-manifolds, one
possibility to approach this might be to introduce boundary
conditions that fix additional structure at an asymptotic infinity,
and to interpret the large diffeomorphisms in terms of the structure
at the infinity\cite{carlip-scat}. The infinity would then be
understood as an ambient physical system. However, for compact
two-manifolds no such reference to an outside system is
possible. The interpretational issue is therefore not at all obvious.

\acknowledgments
We would like to thank Chris Bishop for bringing
Refs.\cite{cowsteg,bishsteg} to our attention, and Abhay Ashtekar,
Martin Bordemann, John Friedman, Don Marolf, and Peter Peld\'{a}n for
discussions. D.\thinspace{}G. is grateful to John Friedman and Karel
Kucha\v{r} for their hospitality during the early stages of this
work.
This work was supported in part by the NSF grant PHY91-05935.

\appendix
\section{A property of the Lebesgue measure}
\label{app:lebesgue}

In this appendix we consider the measure space given by the pair
$(\BbbR,\mu)$, where $\BbbR$ is the real line and $\mu$ its Lebesgue
measure.  We prove that for any given measurable set~$\Delta$, where
$\mu(\Delta)=d>0$, there exists a measurable subset
$\Delta_1\subset\Delta$, so that $0<\mu(\Delta_1)<\delta$, for any
given $\delta<d$. Note that $d$ may be~$\infty$.

To prove this, we cover the real line by  the intervals
$I_n=[n\delta, (n+1)\delta]$, $n\in\BbbZ$. We have
$\mu(\Delta)=\sum_n\mu(\Delta\cap I_n)=d$. Hence there exists
an interval, say~$I_k$, such that $\mu(I_k\cap\Delta)>0$.
We then define $\Delta_1:=\Delta\cap I_k$, which as intersection
of two measurable sets is measurable. It clearly satisfies
$0<\mu(\Delta_1)<\mu(I_k)=\delta$.

\section{$\sltwoz$ action on the circle}
\label{app:Sone}

We have noted that the principal series irreducible unitary
representations ${\tilde \repsym}_s^+$ ($s\in\BbbR$) and ${\tilde
\repsym}_s^-$ ($s\in\BbbR\setminus\{0\}$) of $\sltwor$ on $L^2(S^1)$
restrict to irreducible unitary representations of $\sltwoz$ on
$L^2(S^1)$\cite{cowsteg}. In this appendix we show that the
associated action of $\sltwoz$ on the ``configuration space" $S^1$ of
$L^2(S^1)$ is nowhere properly discontinuous, and hence the quotient
space $S^1/\sltwoz$ is nowhere locally a manifold.
This reintroduces, at the
level of the decomposition~(\ref{hpm-string}), the difficulties
outlined in Section \ref{sec:intro} for constructing a quantum theory
in which the large diffeomorphisms are treated as gauge.

Recall\cite{coxmos,rankin} that $\sltwoz$ is generated by the two
matrices\footnote{We follow the notation of Ref.\cite{coxmos}.}
\begin{mathletters}
\begin{eqnarray}
S &:=& \pmatrix{1 & 1 \cr -1 & 0 \cr}
\ \ ,
\\
T &:=& \pmatrix{0 & 1 \cr -1 & 0 \cr}
\ \ ,
\end{eqnarray}
\end{mathletters}%
whose only independent relations are
\begin{equation}
S^3 = T^2 = -\openone
\ \ .
\end{equation}
Here $\openone$ denotes the two by two identity matrix as before. In
terms of the $\suoneone$ parametrization~(\ref{suoneonepar}),
$S$~corresponds to $\alpha= {1\over2} + i$ and $\beta= {1\over2}$,
and $T$ corresponds to $\alpha=i$ and $\beta=0$.

We are interested in the $\sltwoz$ action on $S^1$ given
by~(\ref{sltwoactionphi}). It will be useful to identify
$S^1$ with the one-point compactified real line,
${\dot \BbbR}=\BbbR \cup \{ \infty \}$, through
\begin{equation}
\tan (\phi/2) = t
\ \ , \ \
t \in {\dot \BbbR}
\ \ ,
\label{dotrpara}
\end{equation}
where $\phi=\pi$ is understood to correspond to $t=\infty$.

Let us denote the action of $\sltwoz$ on ${\dot \BbbR}$ arising
through (\ref{sltwoactionphi}) and~(\ref{dotrpara}) by $t\mapsto
\rinftyrep(M)t$, $M\in\sltwoz$. We have
\begin{mathletters}
\begin{eqnarray}
\rinftyrep(S)t &=& -1/(1 + t)
\ \ ,
\\
\rinftyrep(S^{-1})t &=& -(1+t)/t
\ \ ,
\\
\rinftyrep(T)t &=& \rinftyrep(T^{-1})t = -1/t
\ \ .
\end{eqnarray}
\end{mathletters}%
It is clear that $\rinftyrep(S)$ and $\rinftyrep(T)$ both act on
${\dot \BbbR}$ freely, and $\rinftyrep(S)$ and $\rinftyrep(T)$
respectively generate the sugroups $\Gamma_S := \{ \openone_{\dot
\BbbR}, \rinftyrep(S), \rinftyrep(S^{-1}) \} \simeq \BbbZ_3$ and
$\Gamma_T := \{ \openone_{\dot \BbbR}, \rinftyrep(T) \} \simeq
\BbbZ_2$ of $\Gamma := \{ \rinftyrep(M) \mid M\in \sltwoz \} \simeq
\sltwoz/\{\openone,-\openone\} = \psltwoz$. Here $\openone_{\dot
\BbbR}$ denotes the identity transformation on~${\dot \BbbR}$. We
wish to characterize the quotient space $S^1/\sltwoz \simeq {\dot
\BbbR}/\Gamma$.

Consider first ${\dot \BbbR}/\Gamma_S$. We have
\begin{mathletters}
\begin{eqnarray}
\rinftyrep(S)[0,\infty) &=& [-1,0)
\ \ ,
\\
\rinftyrep(S)[-1,0) &=& [-\infty,-1)
\ \ ,
\\
\rinftyrep(S)[-\infty,-1) &=& [0,\infty)
\ \ .
\end{eqnarray}
\end{mathletters}%
Every equivalence class in ${\dot \BbbR}/\Gamma_S$ has therefore a
unique representative in $[0,\infty)$. This means that we can
identify ${\dot \BbbR}/\Gamma_S \simeq [0,\infty)$, with the circular
topology that identifes $[0,\infty) \simeq S^1$.

Next, consider the action of $\Gamma_T$ on ${\dot \BbbR}/\Gamma_S
\simeq [0,\infty)$. Composing $\rinftyrep(T)$ with $\rinftyrep(S)$ or
$\rinftyrep(S^{-1})$ as appropriate, one finds that this action is
\begin{equation}
t \mapsto \cases{
t/(1-t) \ , &$0\leq t < 1$ \ , \cr
t-1 \ , &$1\leq t < \infty$ \ . \cr}
\label{fmapt}
\end{equation}
For any $t_0 \in [1,\infty)$, iterating (\ref{fmapt}) eventually
gives the fractional part of~$t_0$, $\fracpart (t_0) \in [0,1)$.
Therefore, $S^1/\sltwoz$ can be identified with
the quotient space of $[0,1)$ under iteration of the map
$F: [0,1) \to [0,1)$, defined by
\begin{equation}
F(t) = \fracpart \! \left( {1 \over 1-t} \right)
\ \ .
\end{equation}
All the fixed points of $F$ are unstable, and it is clear that
$S^1/\sltwoz$ is not a manifold.
Note that $F$ is closely related to
the Gauss map, $t\mapsto \fracpart (1/t)$, which is well known in
ergodic theory\cite{billingsley,cornfeld}, and which has also been
encountered in the qualitative analysis of Bianchi type IX
cosmology\cite{barrow,khalaetal}.

\end{document}